\documentclass[prb,aps,twocolumn,floats,amsmath,amssymb]{revtex4}
\usepackage{bm,graphicx,epsf}
\newcommand{\be}{\begin{equation}}
\newcommand{\ee}{\end{equation}}   
\newcommand{\bea}{\begin{eqnarray}}
\newcommand{\eea}{\end{eqnarray}}
\newcommand{\phrl}[1]{Phys.~Rev.~Lett. {\bf #1}}
\newcommand{\phrb}[1]{Phys.~Rev.~B {\bf #1}}
\newcommand{\cmat}[1]{arXiv:cond-mat/#1}
\newcommand{\bib}{\bibitem}
\newcommand{\lb}{\left[}
\newcommand{\rb}{\right]}
\newcommand{\lp}{\left(}
\newcommand{\rp}{\right)}
\newcommand{\lf}{\left\{}
\newcommand{\rf}{\right\}}

\renewcommand{\r}{{\bf r}}

\renewcommand{\k}{{\bf k}}
\newcommand{\R}{{\bf R}}

\begin{document}

\title{Ginzburg-Landau theory of noncentrosymmetric superconductors}
\author{ Soumya P. Mukherjee and Sudhansu S. Mandal}
\affiliation{Theoretical Physics Department, Indian Association for the Cultivation of Science, Jadavpur, Kolkata 700 032, India}

\date{\today}


\begin{abstract}
The data of temperature dependent superfluid density $n_s(T)$
in Li$_2$Pd$_3$B and Li$_2$Pt$_3$B [Yuan {\it et al.}, \phrl97, 017006 (2006)] 
show that a sudden change of the slope of $n_s (T)$ occur
at slightly lower than the critical temperature. Motivated
by this observation, we 
microscopically derive the Ginzburg-Landau (GL) equations for noncentrosymmetric
superconductors with Rashba type spin orbit interaction. Cooper pairing is
assumed to occur between electrons only in the same spin split band and 
pair scattering is allowed to occur between two spin split bands.   
The GL theory of such a system predicts two transition temperatures, the higher
of which is the conventional critical temperature $T_c$
while the lower one $T^*$ corresponds
to the cross-over from a mixed singlet-triplet phase at lower temperatures
to only spin-singlet or spin-triplet 
(depending on the sign of the interband scattering potential)
phase at higher temperatures.
As a consequence, $n_s (T)$ shows a kink at this cross-over temperature.
We attribute the temperature at which sudden change of slope occurs in
the observed $n_s (T)$ to the temperature $T^*$.
This may also be associated with the observed kink in the penetration depth
data of CePt$_3$Si.
We have also estimated critical field near critical temperature.

\end{abstract}

\maketitle

\section{Introduction}

The spin-orbit (SO) coupling of electrons in noncentrosymmetric crystals
lifts the spin degeneracy and hence splits the energy bands. For weak SO coupling,
band splitting energy $E_{SO}$ is smaller than the superconducting energy scales.
In this case, pairing potential may still be chosen as a function of spin
and momentum of quasiparticles near the Fermi surface unaffected by the SO
coupling. \cite{Edelstein,Gorkov,Sigrist}
 In the opposite limit, i.e., when the band splitting energy exceeds
the superconducting critical temperature $T_c$, the electrons with opposite
momenta form Cooper pairs only if they are from same nondegenerate band.
\cite{Curnoe,Samokhin2,Samokhin,Mandal,Mineev1}
Interband pairing in this case can be neglected.
Due to the lack of inversion symmetry in the underlying crystal,
the superconducting order parameter may, in general, be an admixture 
\cite{Gorkov} of spin-singlet and spin-triplet components, i.e., 
the gap function may be decomposed as $\Delta_\k
=[\psi_k \hat{\sigma}_0 + \bm{d}_\k\cdot \hat{\bm{\sigma}}]i\hat{\sigma}_y$,
where $\psi_\k$ is the spin-singlet component and $\bm{d}_\k$ is the
spin-triplet component of the order parameter, and $\sigma$'s are the Pauli
matrices. The spin-triplet component is however
possible only in the presence of spin-triplet channel in the pairing interaction
potential, even in the presence of SO splitting.

The recent discovery \cite{Bauer} of superconductivity in CePt$_3$Si which is 
noncentrosymmetric, has raised interest in the properties of superconductors
without inversion symmetry. A flurry of noncentrosymmetric
heavy fermion compounds like
UIr (Ref.~\onlinecite{Akazawa}), CeRhSi$_3$ (Ref.~\onlinecite{Kimura}), 
CeIRSi$_3$ (Ref.~\onlinecite{Onuki}) exhibiting superconductivity have 
been discovered since then. All of these compounds are strongly correlated: 
Both antiferro magnetism and superconductivity coexist \cite{Bauer}
in CePt$_3$Si, in particular. On the other hand
recently discovered Li$_2$Pd$_3$B (Ref.~\onlinecite{Togano})
and Li$_2$Pt$_3$B (Ref.~\onlinecite{Togano2}) compounds are not of
strongly correlated type and thus may be ideally
used to explore the properties of noncentrosymmetric superconductivity.
 The band structure calculation \cite{Bose}
in CePt$_3$Si reveals that 
 $500K \lesssim E_{SO} \lesssim 2000K$, i.e., $E_{SO}$ is much larger than $T_c$
which is reported to be $0.75K$. Therefore the pairing between electrons in two
different spin split bands can be neglected for CePt$_3$Si and so are in the case
of Li$_2$Pd$_3$B and Li$_2$Pt$_3$B compounds.
In this paper, we consider this assumption.

Both the penetration depth data \cite{Bonalde}
and thermal conductivity data \cite{Behnia} in CePt$_3$Si
seem to suggest the existence of line nodes in the system. However, a theoretical
model \cite{Sigrist2} consisting of
 mixed singlet and triplet order parameters with no line node may also explain  
the penetration depth data \cite{Yuan} at low temperatures. 
This model reasonably fits 
also with the data of superfluid density $n_s(T)$ 
in Li$_2$Pd$_3$B and Li$_2$Pt$_3$B
at low temperatures. However this model alone can not explain the sudden change
in slope of $n_s(T)$ at some characteristic temperature 
that has been clearly observed \cite{Yuan} in these systems, specially
in Li$_2$Pt$_3$B. This motivates us to study Ginzburg-Landau (GL) theory for
two component order parameters associated with two spin split bands formed
in the presence of SO interaction. In this theory, we have considered attractive
intraband pairing potential and attractive or repulsive 
interband pair scattering potential. As a consequence we show, apart from the
conventional superconducting critical temperature, that there is another
characteristic temperature $T^*$ at which superconducting order parameter
undergoes a cross-over from a mixed singlet-triplet phase at lower temperatures
to only triplet or singlet phase at higher temperatures. The superfluid density
shows a kink in its behaviour at the temperature $T^*$.

The article is organized as follows. 
In section~\ref{sec:basics}, we review some important aspects of the Hamiltonian
for a noncentrosymmetric superconductor. It corresponds to two bands with
opposite helicity. 
Following the method of semiclassical gradient expansion \cite{Ting},
we microscopically derive Ginzburg-Landau equations for such a superconductor
in section~\ref{sec:order-parameter}. Both the intraband pairing potential
and interband pair scattering potential have been considered. As a consequence,
the Ginzburg-Landau equations for two bands are coupled.
We analyze the GL equations in terms of the singlet and triplet order parameters
in section~\ref{sec:T_c} by combining GL equations for two separate bands.
We find that the new GL equations are decoupled in the linear order of 
the singlet and triplet order parameters. This predicts two different transition
temperatures: The higher of these corresponds to the usual superconducting
transition temperature and the lower one describes a transition from mixed
singlet-triplet phase at lower temperatures to only triplet or singlet phase,
depending on the sign of the interband pair scattering potential,
at higher temperatures.
We estimate the value of critical magnetic field near $T_c$ in
section~\ref{sec:H_c}.
We finally summarize our results and discuss experimental consequences in
section~\ref{sec:summary}.

\section{Noncentrosymmetric superconductors}
\label{sec:basics}
We begin this section with a brief introduction to the 
model Hamiltonian for noncentrosymmetric superconductors.
The normal state Hamiltonian 
\cite{Edelstein,Gorkov,Sigrist,Curnoe,Samokhin2,Samokhin} for the electrons in 
a band of lattice without inversion symmetry is 
\be
  H_0 = \sum_{\k, s}\xi_\k c_{\k s}^\dagger c_{\k s} + \sum_{\k,s,s'}
       \bm{g}_\k \cdot \bm{\sigma}_{ss'}c_{\k s}^\dagger c_{\k s'} \, ,
\label{H_normal}
\ee
where electrons with momentum $\k$ and spin $s\, (=\uparrow \text{or}\downarrow)$
 are created (annihilated) by the operators $c_{\k s}^\dagger$ ($c_{\k s}$),
$\xi_\k$ is the band energy measured from the Fermi energy $\epsilon_F$. The 
second term in the Hamiltonian(\ref{H_normal}) breaks parity as 
$\bm{g}_{-\k} = -\bm{g}_\k$ for a non-centrosymmetric system. For a system like 
Heavy fermion compound CePt$_3$Si
which has layered structure, $H_0$ is considered to be two-dimensional. For such 
a system of electrons with band mass $m$, $\xi_\k = \frac{\k^2}{2m} -\epsilon_F$ 
and $\bm{g}_\k = \alpha \bm{\eta}_\k$ where $\bm{\eta}_k = \hat{n} \times \k$,
i.e, the spin-orbit interaction is of Rashba type where $\alpha$ is called 
Rashba parameter. Here $\hat{n}$ represents the axis of non-centrosymmetry which 
is perpendicular to
the plane of the system. Due to the breaking down of the parity, spin degeneracy 
of the band is lifted; by diagonalizing $H_0$, one finds two spin-split bands 
with energies $\xi_{\k\lambda} = \xi_k+\lambda \alpha \vert \k\vert$ where 
$\lambda = \pm$ describes helicity of the spin-split bands. Therefore in the
 diagonalized basis $H_0$ (\ref{H_normal}) becomes $H_0 = \sum_{\k , 
\lambda = \pm} \xi_{\k\lambda}\tilde{c}_{\k \lambda}^\dagger\tilde{c}_{\k 
\lambda} $,where $\tilde{c}_{\k \lambda} =\lp c_{\k \uparrow} -  
\lambda \Lambda_\k^\ast 
 c_{\k \downarrow} \rp /\sqrt{2}$ is the electron destruction 
operator and $\tilde{c}_{\k \lambda}^\dagger =\lp c_{\k \uparrow}^\dagger -  
\lambda \Lambda_\k c_{\k \downarrow}^\dagger \rp /\sqrt{2}$ is the 
electron creation operator in band $\lambda$ with momentum $\k$ where
$\Lambda_\k = -i\exp (-i\phi_\k)$ with $\phi_\k$ being the angle of $\k$ 
with $\hat{x}$-axis. The Fermi momenta in these bands are $k_F^\lambda = 
\sqrt{k_F^2+m^2\alpha^2} -\lambda m \alpha$ where $k_F = \sqrt{2m\epsilon_F}$ is 
the Fermi momentum in the absence of band splitting. The density of electronic 
states at Fermi energy in these bands may be found as
$\nu_\lambda = \frac{m}{2\pi}\lp 1 - \lambda m\alpha / \sqrt{k_F^2+m^2\alpha^2} 
\rp$.

Band structure calculation \cite{Bose}
on CePt$_3$Si reveals that the energy difference 
between two spin-split bands near $k_F$ is 50--200 meV which is much larger than 
the superconducting critical temperature, $k_BT_c \approx 0.06$ meV. The 
formation of Cooper pairing between electrons in different spin-split bands may 
thus be ignored \cite{Curnoe,Samokhin2,Samokhin,Mandal},
 i.e., $\langle \tilde{c}_{\k\lambda} \tilde{c}_{-\k\lambda'}
\rangle $ is finite only when $\lambda' = \lambda$. However the scattering of 
pairs between two spin-split bands are allowed. 
The Hamiltonian for the system may then be written as
\be
\label{H_tot}
  H =\sum_{\k ,\lambda=\pm}\xi_{\k\lambda}\tilde{c}_{\k\lambda}^\dagger
\tilde{c}_{\k \lambda}
+\sum_{\k,\k'}\sum_{\lambda ,\lambda'}
        V_{\lambda\lambda^{'}}{(\k,\k')}\tilde{c}_{\k \lambda}^\dagger
       \tilde{c}_{-\k \lambda}^\dagger \tilde{c}_{-\k' \lambda^{'}}
  \tilde{c}_{\k' \lambda^{'}}
\ee
where $V_{\lambda\lambda'}(\k ,\k')$ represents intraband pair potential 
and interband pair scattering potential.

\section{Ginzburg-Landau Equations}
\label{sec:order-parameter}

The total second quantized Hamiltonian in the real space can 
be written by performing a Fourier transformation of equation (\ref{H_tot}). 
It then takes the form
\bea
\label{H_real space}
 &&{\cal H}=\int d\r  \varphi^\dagger_{\lambda} (\r)
 \lp {\frac{(\bm{p}+e\bm{A})^2}{2m}+\mu}\rp \varphi_{\lambda}(\r)+\nonumber\\&& 
\int\int d\r d\r'\varphi_\lambda^\dagger(\r)  
 \varphi_\lambda^\dagger(\r')V_{\lambda \lambda{'}}(\r-\r')
 \varphi_{\lambda^{'}}(\r')\varphi_{\lambda^{'}}(\r)
\eea
Here $\varphi_{\lambda}(r)$ is the 
field operator for electrons in band $\lambda$ at position $\r$ and 
the repeated indices's denotes summation. $V_{\lambda\lambda'} (\r-\r')$ 
denotes intraband pairing as well as inter-band pair scattering potential
The vector potential $\bm{A}$ which preserves gauge invariance is introduced.
From here after we consider unit system: $\hbar=1$, $k_B=1$ and $c=1$.   
In Gor'kov's weak coupling theory, the equation of motion of the normal 
and anomalous Green's functions in each spin-split band can be written as
\bea
\label{G}
&&\lp{i\omega_n-\frac{(\bm{p}+e\bm{A})^2}{2m}+\mu}\rp {\cal G}_{\lambda}{(\r,\r';\omega_n)}
  \nonumber\\&& +\int d\r''\Delta_\lambda{(\r,\r')}{\cal F}_\lambda^\dagger
{(\r'',\r';\omega_n)}=\delta{(\r-\r')} \\
\label{F}
&&\lp{-i\omega_n-\frac{(\bm{p}-e\bm{A})^2}{2m}+\mu}\rp{\cal F}_\lambda{(\r,\r';\omega_n)}
 \nonumber\\&&-{\int d\r'' \Delta_\lambda^\ast(\r,\r'){\cal G}_\lambda
{(\r'',\r';\omega_n)}}=0,
\eea 
where ${\cal G_\lambda}{(\r,\r';{\omega}_n)}$
and ${\cal F_\lambda}{(\r,\r';{\omega}_n)}$ respectively
are normal and
anomalous quasiparticle Green's functions in band $\lambda$, and 
$\Delta_\lambda^\ast(\r,\r')$is the gap function which can be written as
\be
\label{delta}
{\Delta_\lambda^\ast(\r,\r')}=-T \sum_{n,\lambda'}V_{\lambda\lambda'}(\r,\r')
{\cal F}_{\lambda'}^{\dagger}(\r,\r';\omega_n)
\ee 
where $\omega_n=(2n+1)\pi T$ is the fermionic Matsubara frequency
at temperature $T$. 
Normal state electronic Green's function 
$\hat G_{\lambda}(\r,\r';\omega_n)$ satisfies the equation
\be 
\label{G_0}
           \lp{i\omega_n-\frac{(\bm{p}+e\bm{A})^2}{2m}+\mu}\rp G_{\lambda}
(\r,\r';\omega_n)=\delta(\r-\r')
\ee
In terms of $G_{\lambda}(\r,\r';\omega_n)$, a self consistent solution 
of Eqs.~(\ref{G}) and (\ref{F}) becomes
\begin{widetext}
\bea
\label{G_new}
{\cal G}_\lambda (\r,\r';\omega_n) &=& G_{ \lambda}(\r,\r';\omega_n)- 
\int d\r_{1}d\r_{2} G_{ \lambda}(\r,\r_{1};\omega_n)
 \Delta_\lambda(\r_{1},\r_{2}){\cal F}_\lambda^\dagger(\r_{2},\r';\omega_n) \\ 
 {\cal F}_\lambda^\dagger (\r,\r';\omega_n) &=& \int d\r_{1}d\r_{2}
     G_{ \lambda}(\r,\r_{1};-\omega_n)\Delta_{\lambda}^\ast(\r_{1},\r_{2})
  \hat{\cal G}_{\lambda}(\r_{2},\r';\omega_n) 
\eea
In the absence of $\bm{A}$, the normal state Green's function is translationaly 
invariant and can be written in momentum space as 
${\widetilde G}_{\lambda}=1/(i\omega_n-\xi_{k\lambda})$. 
In a semiclassical approximation \cite{Ting}, 
the role of $\bm{A}$ is to generate a
phase in the single particle normal state Green's function:
\be
\label{spa} 
 G_{\lambda}(\r,\r';\omega_n)={\widetilde G}_{\lambda}(\r,\r';\omega_n) 
            \exp\lp{-ie\int_{\r'}^{\r}{\bf d\bm{s}\cdot \bm{A}(\bm{s})}}\rp
\ee
where the integration is over a straight line path from $\r'$ to $\r$. 
Close to the superconducting transition temperature, magnitude of 
order parameter is small and its smallness allows us to expand 
${\cal F}^{\dagger}$ and ${\cal G}$ in terms of it for each individual spin 
split band:
\bea
\label{G_exp}
 & &{\cal G}_\lambda (\r,\r';\omega_n) = G_{ \lambda}(\r,\r';\omega_n)-
 \int d\r_{1}d\r_{2}
 G_{ \lambda}(\r,\r_{1};\omega_n) \Delta_\lambda(\r_{1},\r_{2})
  \int d\r_{3}d\r_{4}
G_{ \lambda}(\r_{2},\r_{3};-\omega_n)\Delta_\lambda^{\ast}(\r_{3},\r_{4})
 G_{\lambda}(\r_{4},\r';\omega_n) \\
\label{F_exp}
& & {\cal F}_\lambda^\dagger (\r,\r';\omega_n) = \int d\r_{1}d\r_{2}
   G_{ \lambda}(\r,\r_{1};-\omega_n)\Delta_{\lambda}^\ast(\r_{1},\r_{2}) \nonumber\\
& &  \times
\lb G_{ \lambda}(\r_{2},\r';\omega_n)-\int d\r_{3}d\r_{4}d\r_{5}d\r_{6}
 G_{ \lambda}(\r_{2},\r_{3};\omega_n)
 \Delta_{\lambda}(\r_{3},\r_{4})G_{ \lambda}(\r_{4},\r_{5};-\omega_n)
\Delta_\lambda^{\ast}(\r_{5},\r_{6})
G_{ \lambda}(\r_{6},\r';\omega_n)\rb 
\eea
Substituting Eq.~(\ref{F_exp}) in Eq.(\ref{delta}) and writing
\be
\label{delta_sum}
 \Delta_{\lambda}^{\ast}(\r,\r')=\Delta_{\lambda_I}^{\ast}(\r,\r')+
\Delta_{\lambda_{II}}^{\ast}(\r,\r')
\ee
we find
\bea
\label{delta_1}
  \Delta_{\lambda_I}^{\ast}(\r,\r') &=&
-T\sum_{n,\lambda'}V_{\lambda \lambda'}(\r,\r')
\int d\r_{1}d\r_{2}G_{ \lambda'}(\r,\r_{1};-\omega_n)
\Delta_{\lambda'}^\ast(\r_{1},\r_{2})G_{ \lambda'}(\r_{2},\r';\omega_n), \\ 
\label{delta_2}
 \Delta_{\lambda_{II}}^{\ast}(\r,\r') &=& T\sum_{n,\lambda'}
V_{\lambda \lambda'}(\r,\r')\int d\r_{1-6} 
G_{\lambda'}(\r,\r_{1};-\omega_n)\Delta_{\lambda'}^\ast(\r_{1},\r_{2})
G_{\lambda'}(\r_{2},\r_{3};\omega_n)
\Delta_{\lambda'}(\r_{3},\r_{4}) \nonumber \\
&\times &  G_{ \lambda}(\r_{4},\r_{5};-\omega_n)
\Delta_{\lambda'}^{\ast}(\r_{5},\r_{6})
G_{ \lambda'}(\r_{6},\r';\omega_n)
\eea
Expressing the order parameter $\Delta_{\lambda}^\ast(\r_{1},\r_{2})$ in terms
of center of mass coordinate $\bm{R} = (\r_1 +\r_2)/2$ of the pair
and relative coordinate 
$\bm{\rho} = \r_1 -\r_2$ of the pair and 
making Fourier transform with respect to the 
relative coordinate, we can express $\Delta_{\lambda_I}^\ast$ in 
Eq.~(\ref{delta_1}) as the sum of two terms:
\be
\Delta_{\lambda_I}^\ast =  \Delta_{\lambda_{Ic}}^\ast
+ \Delta_{\lambda_{Ig}}^\ast
\label{delta_I}
\ee
where
\be
\label{del_Ic}
\Delta_{\lambda_{Ic}}^{\ast}(\R,\k)=-T\sum_{n,\lambda'}
\int\frac{d^{2}\k'}{(2\pi)^2}
V_{\lambda\lambda'}(\k-\k')\frac{1}{\omega_{n}^2+
\xi_{\k'\lambda'}^2}\Delta_{\lambda'}^{\ast}(\R,\k'),
\ee
\be
\label{del_Ig}
\Delta_{\lambda_{Ig}}^{\ast}(\R,\k)=-T\sum_{n,\lambda'}\int\frac{d^{2}\k'}{2{(2\pi})^2}
V_{\lambda\lambda'}(\k-\k')\lf\frac {1}{(2m)^2} \frac{2\xi_{\k'\lambda'}^2
-6\omega_{n}^2}{{(\omega_{n}^2+\xi_{\k'\lambda'}^2)}^3}{{(k'_{x}\Pi_{x}+k'_{y}
\Pi_{y})}^2}-\frac {1}{2m}\frac{\xi_{\k'}{\bf\Pi}^2}{{(\omega_{n}^2+
\xi_{\k'\lambda'}^2)}^2}\rf \Delta_{\lambda'}^{\ast}(\R,\k')
\ee
Similarly we find from Eq.~(\ref{delta_2}),
\be
\label{delta_II}
\Delta_{\lambda_{II}}^\ast(\R,\k)=T\sum_{n,\lambda'}\int
\frac{d^{2}\k'}{(2\pi)^2}V_{\lambda\lambda'}(\k-\k')
\frac{1}{{(\omega_{n}^2+\xi_{\k'\lambda'}^2)}^2}
 {|\Delta_{\lambda'}(\R,\k')|}^2\Delta_{\lambda'}^{\ast}(\R,\k')
\ee 
\end{widetext}

We assume the interaction potential to be 
\be
\label{V_real}
V_{\lambda\lambda'}(\k-\k')=
 -V_{\lambda\lambda'}\hat{k} \cdot \hat{k}'
= -V_{\lambda\lambda'}\lb\Lambda_\k^{\ast}\Lambda_{\k'}
   +\Lambda_\k\Lambda_{\k'}^{\ast}\rb  
\ee
where interaction strength $V_{\lambda\lambda'} > 0 $ for $\lambda = \lambda'$
and it may have either sign when $\lambda \neq \lambda'$. The potential
$V_1 = -V_{\lambda\lambda'}\Lambda_\k^{\ast}\Lambda_{\k'}$ leads to the
order parameter $\Delta_{\lambda ,1} \Lambda_\k$ which in turn corresponds 
to $s$-wave pairing in singlet channel, and $p$-waves 
for spin up-up and down-down triplet channels.
The other part of the potential (\ref{V_real}), 
$V_2 = -V_{\lambda\lambda'}\Lambda_\k \Lambda_{\k'}^\ast$
will help to induce order parameter $\Delta_{\lambda ,2} \Lambda_\k^\ast$.
This new order parameter corresponds to $d$-wave in singlet channel,
and $p$-wave and $f$-wave for spin up-up and down-down triplet channels
respectively.
Thus we can write the new form of the order parameter as,
\be
\label{new_delta}
\Delta_{\lambda}^{\ast}(\R,\k)=\Delta_{\lambda,1}^{\ast}(\R)\Lambda_{k}^{\ast}
+\Delta_{\lambda,2}^{\ast}(\R)\Lambda_{k}.
\ee
Inserting the form of $V_{\lambda\lambda'}(\k -\k')$ in Eq.~(\ref{V_real}) and 
$\Delta_{\lambda}^{\ast}(\R,\k)$ in Eq.~(\ref{new_delta}) into Eqs.~(\ref{del_Ic}
-- \ref{delta_II}), we find
\begin{widetext}
\bea
\label{real_del_1c}
\Delta_{\lambda_{Ic}}^{\ast}(\R,\k) &=&\ln \lp\frac {2{e^\gamma}\omega_D}{\pi T}\rp
\sum_{\lambda'}g_{\lambda\lambda'}\lp \Delta_{\lambda',1}^{\ast} \Lambda_{k}^\ast\ +
\Delta_{\lambda',2}^{\ast} \Lambda_{k}\rp \\
\label{real_del_1g}
\Delta_{\lambda_{Ig}}^{\ast}(\R,\k) &=& -\frac{\alpha}{8}
 \sum_{\lambda'}g_{\lambda\lambda'}v_{F\lambda'}^2
\lb \lp 2\bm{\Pi}^2\Delta_{\lambda',1}^{\ast}
+\bm{\Pi}_{-}^2\Delta_{\lambda',2}^{\ast}\rp \Lambda_{k}^\ast 
+\lp 2\bm{\Pi}^2\Delta_{\lambda',2}^{\ast} 
+\bm{\Pi}_{+}^2\Delta_{\lambda',1}^{\ast}
\rp \Lambda_{k}\rb \\
\label{real_del_2}
 \Delta_{\lambda_{II}}^{\ast}(\R,\k) &=& -\alpha\sum_{\lambda'}g_{\lambda\lambda'}
 [\lp|\Delta_{\lambda',1}|^2+2|\Delta_{\lambda',2}|^2|\rp
\Delta_{\lambda',1}^{\ast}\Lambda_{\k}^\ast 
+\lp 2|\Delta_{\lambda',1}|^2 +|\Delta_{\lambda' ,2}|^2\rp 
\Delta_{\lambda',2}^{\ast}\Lambda_{\k}]
\eea
where dimensionless interaction strength
$g_{\lambda\lambda'}=\frac{1}{2} V_{\lambda\lambda'}\nu_{\lambda'}$, 
$\gamma=0.5772$ is the Euler constant, $\omega_D$ is the Debye frequency, 
$v_{F\lambda}$ is the Fermi velocity for band $\lambda$ and 
$\alpha=\frac{7\zeta(3)}{8{(\pi T)}^2}$.
Further $\bm{\Pi}=-i\bm{\nabla}_{\R}-2e\bm{ A}(\R)$ and ${\Pi}_{\pm}=
\Pi_{x}\pm i\Pi_{y}$. 

Summing expressions (\ref{real_del_1c}--\ref{real_del_2}) and equating the
sum with Eq.~(\ref{new_delta}) and then by comparing
coefficients of $\Lambda_\k^\ast$ and $\Lambda_\k$ we find the GL 
equations for each band with primary as well as induced order parameters: 
\be
\label{delta_lambda1}
\Delta_{\lambda,1}^{\ast}(\R)=\ln \lp \frac {2{e^\gamma}\omega_D}{\pi T}\rp
\sum_{\lambda'}g_{\lambda\lambda'}\Delta_{\lambda',1}^{\ast}-
\frac{\alpha}{8}\sum_{\lambda'}g_{\lambda\lambda'}v_{F\lambda'}^2  \lp 
2\bm{\Pi}^2\Delta_{\lambda',1}^{\ast}+\Pi_{-}^2\Delta_{\lambda',2}^{\ast}\rp  
 -\alpha\sum_{\lambda'}g_{\lambda\lambda'}\lp|\Delta_{\lambda',1}|^2+
2|\Delta_{\lambda',2}|^2 \rp \Delta_{\lambda',1}^{\ast}
\ee
\be
\label{delta_lambda2}
\Delta_{\lambda,2}^{\ast}(\R)=\ln \lp \frac {2{e^\gamma}\omega_D}{\pi T}\rp
\sum_{\lambda'}g_{\lambda\lambda'}\Delta_{\lambda',2}^{\ast}-
\frac{\alpha}{8}\sum_{\lambda'}g_{\lambda\lambda'}v_{F\lambda'}^2  \lp 
2\bm{\Pi}^2\Delta_{\lambda',2}^{\ast}+\Pi_{+}^2\Delta_{\lambda',1}^{\ast}\rp
 -\alpha\sum_{\lambda'}g_{\lambda\lambda'}\lp|\Delta_{\lambda',2}|^2+
2|\Delta_{\lambda',1}|^2 \rp \Delta_{\lambda',2}^{\ast}
\ee
Note that gradient of $\Delta_{\lambda,1}^{\ast}(\R)$ leads to the
induction of $\Delta_{\lambda,2}^{\ast}(\R)$. A self consistent solution of
these order parameters involve simultaneous solution 
of Eqs.~(\ref{delta_lambda1}) and (\ref{delta_lambda2}).
The transition temperature $T_c$ however may be obtained from the linear in
$\Delta_{\lambda,1}^{\ast}(\R)$ terms in Eq.~(\ref{delta_lambda1}). 
Solving the matrix equation, one finds 
\be
T_c = \lp\frac{2e^\gamma \omega_D}{\pi}\rp \exp \lb -\frac{1}{g_2}\rb \, ; \,
g_{1,2} = \frac{1}{2}\lb (g_{++}+g_{--}) \pm 
\sqrt{(g_{++}-g_{--})^2+4g_{+-}g_{-+}}\rb
\ee
The critical temperature should be
 determined by the solution min$(g_1,\, g_2)$, i.e., $g_2$ in
contrary to the consideration of Ref.~\onlinecite{Mineev1}.
The other solution $g_1$ does not have any physical importance.
However in a certain physical situation as we discuss in
the next section, this redundant solution
gets renormalized to a value less than $g_2$ and manifests itself to
a physical solution.
We choose a special situation when $g_{++} = g_{--}$ and
$g_{+-} = g_{-+}$, i.e., the intra as well as inter band strengths
 of interaction are independent of bands although they are different from
each other in general. This assumption is reasonable since $g_{\lambda\lambda'}$
is dimensionless and is the product of $V_{\lambda\lambda'}$ and 
$\nu_{\lambda'}$, i.e., a density of states weighted interaction
strength. The matrix $\hat{g}$ is positive definite, i.e.,
$g_{++}>0$, $g_{--} >0$, and $det(\hat{g}) >0$. This indicates
$g_{+-}$ may have either of the signs. By this choice,
\be
T_c = \lp\frac{2e^\gamma \omega_D}{\pi}\rp \exp \lb -\frac{1}{g_{++}-\vert g_{+-}
\vert } \rb
\label{TC}
\ee

\section{Transition Temperatures}
\label{sec:T_c}


Order parameters $\Delta_{\lambda ,1}$ and $\Delta_{\lambda ,2}$ consist of
both singlet and triplet components: they are $\Delta_{s,l} = 
(\Delta_{+ ,l} - \Delta_{- ,l})/2$ and $ \Delta_{t,l} = (\Delta_{+ ,l} + 
\Delta_{- ,l})/2$ respectively \cite{Mandal}, where $l = 1$ or $2$. We thus find 
the GL equations for $\Delta_{s,1}$ and $\Delta_{t,1}$ derivable from 
Eq.~(\ref{delta_lambda1}) as
\bea
\label{del_t}
&&\lp 1-\tilde g_{++}-\tilde g_{+-}\rp \Delta_{t,1}^\ast(\R)+\frac{\alpha}{16} 
\lp g_{++}+g_{+-}\rp\lb v_{F,1}^2 \lp 2 \bm{\Pi}^2\Delta_{t,1}^\ast(\R)+
\Pi_{-}^2 \Delta_{t,2}^\ast(\R)\rp - v_{F,2}^2 \lp 2\bm{\Pi}^2
\Delta_{s,1}^\ast(\R)+\Pi_{-}^2 \Delta_{s,2}^\ast(\R) \rp \rb
\nonumber\\&& + 
\alpha \lp g_{++}+g_{+-}\rp  \lb \lp |\Delta_{t,1}(\R)|^2 +2|\Delta_{s,1}(\R)|^2
+2|\Delta_{t,2}(\R)|^2+2|\Delta_{s,2}(\R)|^2 \rp \Delta_{t,1}^\ast(\R)
+{\Delta_{s,1}^\ast}^2(\R)\Delta_{t,1}(\R) \right.  \nonumber\\
&& \left.  
+2\Delta_{s,1}^\ast(\R)\Delta_{s,2}^\ast(\R)\Delta_{t,2}(\R)
+2\Delta_{s,1}^\ast(\R)\Delta_{s,2}(\R)\Delta_{t,2}^\ast(\R) \rb=0
\eea
\bea
\label{del_s}
&&\lp 1-\tilde g_{++}+\tilde g_{+-}\rp \Delta_{s,1}^\ast(\R)+\frac{\alpha}{16} 
\lp g_{++}-g_{+-}\rp\lb v_{F,1}^2 \lp 2 \bm{\Pi}^2\Delta_{s,1}^\ast(\R)+
\Pi_{-}^2 \Delta_{s,2}^\ast(\R)\rp - v_{F,2}^2 \lp 2\bm{\Pi}^2
\Delta_{t,1}^\ast(\R)+\Pi_{-}^2 \Delta_{t,2}^\ast(\R) \rp \rb
\nonumber\\&& + 
\alpha \lp g_{++}-g_{+-}\rp  \lb \lp |\Delta_{s,1}(\R)|^2 +2|\Delta_{t,1}(\R)|^2
+2|\Delta_{s,2}(\R)|^2+2|\Delta_{t,2}(\R)|^2 \rp \Delta_{s,1}^\ast(\R)
+{\Delta_{t,1}^\ast}^2(\R)\Delta_{s,1}(\R) \right.  \nonumber\\
&& \left.  
+2\Delta_{t,1}^\ast(\R)\Delta_{t,2}^\ast(\R)\Delta_{s,2}(\R)
+2\Delta_{t,1}^\ast(\R)\Delta_{t,2}(\R)\Delta_{s,2}^\ast(\R) \rb=0
\eea
\end{widetext}
where $ v_{F,1}^2=v_{F_{+}}^2+v_{F_{-}}^2$ and 
$v_{F,2}^2=v_{F_{-}}^2-v_{F_{+}}^2$.
These equations have been written under the assumption that $g_{++}=g_{--}$ 
i.e., the dimensionless intraband interaction strength is independent of 
spin split band and $g_{+-}=g_{-+}$ which is rather obvious. We also define 
$\tilde g_{\lambda\lambda'}=\ln \lp \frac {2{e^\gamma}\omega_D}{\pi T}\rp 
g_{\lambda\lambda'}$.
Similarly we can use Eq.~(\ref{delta_lambda2}) to obtain the GL equations 
for other two order parameters 
$\Delta_{s,2}$ and $\Delta_{t,2}$, which may be obtained 
by making the replacements 
$\Delta_{s,1}\leftrightarrow\Delta_{s,2}$, $\Delta_{t,1}\leftrightarrow
\Delta_{t,2}$ and $\Pi_{-}\leftrightarrow\Pi_{+}$ in Eqs.~(\ref{del_t}) and
(\ref{del_s}).

Equations (\ref{del_t}) and (\ref{del_s}) clearly show the decoupling of
order parameters $\Delta_{t,1}$ and $\Delta_{s,1}$ in their linear order and
as their coefficients are unequal, they have two different critical 
temperatures. The higher one of these two corresponds to the standard critical 
temperature $T_c$
and the lower one corresponds to the temperature at which
the spin-nature of the order parameter changes. The information of this
new transition temperature is however 
hidden in the Eq.~(\ref{delta_lambda1}) as the
GL equations for $\Delta_{+,1}$ and $\Delta_{-,1}$ are coupled in their
linear order. We also observe from the other two GL equations for $\Delta_{s,2}$ 
and $\Delta_{t,2}$ that the transition temperature for both the singlet order 
parameters are same and this is also the case for the two triplet 
order parameters. 
Relative magnitude of singlet transition temperature $T_s$
and triplet transition temperature $T_t$ depends on the sign of interband
interaction $g_{+-}$.

Assuming $g_{+-} <0$, one finds 
\be
T_{t}=\lp \frac {2{e^\gamma}\omega_D} {\pi} \rp\exp\lb-\frac{1}{g_{++}+g_{+-}}\rb 
\ee
by equating the coefficient of $\Delta_{t,1}$ in Eq.~(\ref{del_t}) with zero
at $T_{t}$ which is identified as $T_c$ (\ref{TC}). We now look for existence
of any other characteristic temperature which could be less than $T_c$.
Assuming further that $\Delta_{s,1} = 0$ and $\Delta_{t,2} 
\ll \Delta_{t,1}$ near $T_t$, we find superfluid density which is entirely 
due to triplet order parameter, to be 
\be
n_s \equiv \vert \Delta_{t,1}\vert^2 = -\frac{1}{\alpha}\ln \lp \frac{T}{T_t}\rp
\ee
The coefficient of $\Delta_{s,1}^\ast$ in Eq.~(\ref{del_s}) is now
$1-\tilde{g}_{++}+\tilde{g}_{+-} +3\alpha (g_{++}-g_{+-})
\vert \Delta_{t,1}\vert^2$. Equating it to
be zero at $T=T_s$, we find
\be
 T_{s}=\lp \frac {2{e^\gamma}\omega_D}
{\pi} \rp \exp\lb-\frac{g_{++}-2g_{+-}}{ g_{++}^2-g_{+-}^2}\rb \, .
\ee
and hence $T_s/T_t = \exp [g_{+-}/(g_{++}^2-g_{+-}^2)] < 1$.
The predicted $T_s$ is then the cross-over temperature $T^*$ below 
which both singlet and triplet pairing exist and above which only triplet 
pairing exists. 

\begin{figure}
\vspace{-3cm}
\centerline{\epsfysize=14cm\epsfbox{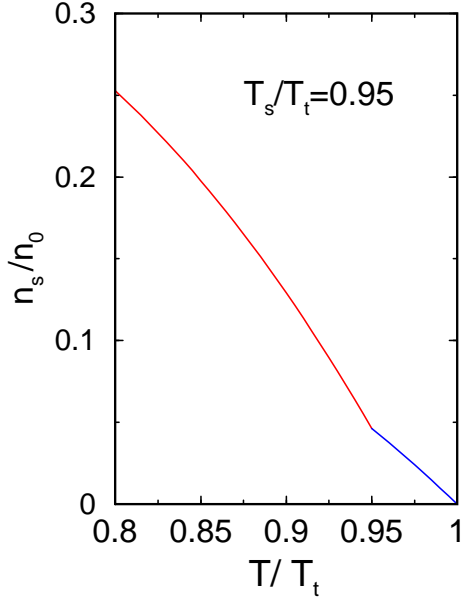}}
\vspace{-3cm}
\caption{Super fluid density $n_s$ in the units of $n_0 = (8\pi^2T_t^2)/(7\zeta
(3))$ as function of $T/T_t$ for $T_{s}/T_{t}=0.95$. $T_t$ is identified as
critical temperature $T_c$ and $T_s$ is identified as cross-over temperature
$T^*$ at which spin symmetry of the order parameter changes.}
\label{fig:n_s}
\end{figure}

Assuming $\Delta_{s,2} \ll \Delta_{s,1}$, we find $\vert \Delta_{s,1}\vert^2
= -\frac{1}{\alpha}\ln \lp \frac{T}{T_s}\rp$ near $T_s$. Therefore total
superfluid density at $T<T_s$,
\be
n_s = \vert \Delta_{s,1}\vert^2 +\vert \Delta_{t,1}\vert^2
= -\frac{1}{\alpha} \lb \ln \lp \frac{T}{T_s}\rp 
+\ln \lp \frac{T}{T_t}\rp \rb
\ee
Figure \ref{fig:n_s} shows the variation of $n_s$ with temperature below
$T_t$ and around $T_s$. It shows a kink at $T=T_s$. 

For attractive interband scattering potential, $g_{+-} >0$ and
 hence $T_c$ coincides with 
\be
T_{s}=\lp \frac {2{e^\gamma}\omega_D} {\pi}\rp  \exp\lb -\frac{1}{g_{++}-g_{+-}}\rb 
\ee
and
\be
 T_{t}= \lp \frac {2{e^\gamma}\omega_D}
{\pi} \rp \exp\lb-\frac{g_{++}+2g_{+-}}{ g_{++}^2-g_{+-}^2}\rb \, .
\ee
becomes cross-over temperature $T^*$ above which the order 
parameter is fully singlet. In this case, $T_t/T_s = 
\exp [-g_{+-}/(g_{++}^2-g_{+-}^2)] < 1$.
Above and below $T^*$, superfluid density is then found to be
$ -(1/\alpha)\ln \lp T/T_s \rp$ and $-(1/\alpha)\lb \ln \lp T/T_s \rp
+\ln \lp T/T_s \rp \rb $ respectively.

\section{The Upper Critical Field}
\label{sec:H_c}

We here estimate the upper critical field near $T=T_{t}>T_ s$.
If the applied magnetic field is along negative 
z-axis, then a convenient gauge choice gives ${\bf A}=(0,-Hx,0)$. To simplify 
the problem by retaining all the essential physics we may consider the 
linearized coupled GL equations for $\Delta_{t,1}$ and $\Delta_{t,2}$.  
We thus find GL equation for $\Delta_{t,1}$ from Eq.~(\ref{del_t}) as
\be
\label{Hc_t}
\ln\lp\frac{T}{T_{t}}\rp \Delta_{t,1}^\ast(\R)+\frac{v_{F,1}^2\alpha}{16}  
\lp 2{\bm{\Pi}^2}\Delta_{t,1}^\ast(\R)+\Pi_{-}^2 \Delta_{t,2}^\ast(\R)\rp=0 
\ee
and similarly for $\Delta_{t,2}$, it is given by
\be
\label{Hc_t'}
\ln\lp\frac{T}{T_{t}}\rp \Delta_{t,2}^\ast(\R)+\frac{v_{F,1}^2\alpha}{16}  
\lp 2{\bm{\Pi}^2}\Delta_{t,2}^\ast(\R)+\Pi_{+}^2 \Delta_{t,1}^\ast(\R)\rp=0 
\ee
By defining $\widetilde \Pi_{\pm}=\frac{\Pi{\pm}}{2\sqrt{eH}}$, it is
easy to show that  
$\lb\widetilde \Pi_{+},\widetilde \Pi_{-}\rb=1$.  
Therefore $\widetilde\Pi_{\pm}$ are regarded as the 
creation and annihilation operators respectively in occupation number space 
such that $\widetilde\Pi_{+}|n>=\sqrt{n+1} |n+1>$ and 
$\widetilde\Pi_{-}|n>=\sqrt{n} |n-1>$ where $|n>$ represents $n$-th Landau
level. Equations (\ref{Hc_t}) and (\ref{Hc_t'}) suggest that the characteristic
order parameter $\Delta_0 = 1/\sqrt{\alpha}$ and the coherence length
$\xi_0 = \sqrt{\frac{v_{F,1}^2\alpha}{8}}$.
The dimensionless order parameters $\psi_{t,j}=
\frac{\Delta_{t,j}}{\Delta{0}},\, (j=1,2)$ 
are then may be expressed as a linear combination of Landau levels: 
 $\psi_{t,j}\ast=\sum_{n=0}^{\infty}{a_{n}}^{t,j}|n>$.  
Therefore Eqs.~(\ref{Hc_t}) and (\ref{Hc_t'}) become
\bea
\label{lan_t}
&& \sum_{n=0}^\infty 2(2n+1)a_{n}^{t,1} |n> + \sqrt{n(n-1)} a_{n}^{t,2}|n-2>\nonumber\\&&= 
   \frac{1}{KeH}\ln\lp\frac{T_{t}}{T}\rp \sum_{n=0}^\infty a_{n}^{t,1} |n>
\eea
\bea
\label{lan_t'}
&& \sum_{n=0}^\infty 2(2n+1)a_{n}^{t,2} |n> + \sqrt{(n+1)(n+2)} a_{n}^{t,1}|n+2> \nonumber\\&& =
\frac{1}{KeH} \ln\lp\frac{T_{t}}{T}\rp \sum_{n=0}^\infty a_{n}^{t,2} |n>
\eea
Equating the coefficients of the lowest Landau level $|0>$ from Eq.~(\ref{lan_t})
we find 
\be
2a_0^{t,1} + \sqrt{2}a_2^{t,2} = \frac{1}{KeH} \ln\lp\frac{T_{t}}{T}\rp a_0^{t,1}
\label{hc1}
\ee
which is one of the equations satisfied by $a_0^{t,1}$ and $a_2^{t,2}$. The other
equation satisfied by these variables is given by
\be
5a_2^{t,2} + \sqrt{2}a_0^{t,1} = \frac{1}{KeH} \ln\lp\frac{T_{t}}{T}\rp a_2^{t,2}
\label{hc2}
\ee
derivable from Eq.~(\ref{lan_t'}).
The solution of the coupled Eqs.~(\ref{hc1}) and (\ref{hc2}) corresponding to
a linear combination of $a_0^{t,1}$ and $a_2^{t,2}$ with the major sharing from
the former leads to the critical field
\be
\label{cr_field}
H_{c2} = \frac{2\sqrt{2}}{3(\sqrt{2}-1)} \frac{1}{e\alpha(v_{F+}^2+v_{F-}^2)}
\ln \lp\frac{T_{t}}{T}\rp 
\ee
near critical temperature $T_c = T_t$.

\section{Summary and Discussion }

\label{sec:summary}

We have analyzed the critical and cross-over temperatures using
equations for order parameters comprising of $\Delta_{s,1}$ and $\Delta_{t,1}$
and neglecting the order parameters $\Delta_{s,2}$ and $\Delta_{t,2}$. This
consideration implies spherically symmetric s-wave in the singlet channel and
the triplet channels are of $p$-waves which have point nodes. On the
other hand, the experiments \cite{Bauer,Behnia,Bonalde,Yuan}
seem to suggest that most of these superconductors,
excepting \cite{Yuan} Li$_2$Pd$_3$B, have lines of nodes. For such a case, equations for 
$\Delta_{s,2}$ and $\Delta_{t,2}$ should be considered and we find that 
the transition and cross-over temperatures remain unaltered.

We observe from the data \cite{Yuan}
of temperature dependent super-fluid density $n_s(T)$
that its slope changes suddenly at $T\sim 0.9T_c$ for Li$_2$Pt$_3$B. This
observation is not however prominent in Li$_2$Pd$_3$B. 
Since the mixed 
singlet-triplet phase of Li$_2$Pd$_3$B has very large singlet component
compared to the triplet component \cite{Yuan}, 
the sudden change in slope of $n_s(T)$ is
invisible at the cross-over temperature. On the other hand, Li$_2$Pt$_3$B
has comparable amount of singlet and triplet components in the mixed 
singlet-triplet phase and thus the cross-over temperature is prominent.

The Knight shift measurements \cite{Zheng}
in Li$_2$Pd$_3$B and Li$_2$Pt$_3$B did not
show any cross-over temperature whatsoever; the former (latter) shows 
singlet (triplet) type of data at all temperatures. However, the error bars
in these data are huge to conclude this subtle effect. 
Moreover, we have not considered the effect of impurity which will
smoothen this cross-over.
The Knight shift measurement in CePt$_3$Si by Yogi {\it et al.} 
\cite{Yogi} seems
to suggest the cross-over temperature is around $0.4K$, from the point of
view of optimistic observation for obvious reason.
A more accurate Knight shift measurement in relatively pure systems will 
directly show the cross-over temperature predicted in this paper. 
Further observed anomaly \cite{Scheidt,Kim} 
in specific heat data of CePt$_3$Si may also be
related with this cross-over temperature.

To summarize,
we have microscopically derived the Ginzburg-Landau equations for a 
noncentrosymmetric superconductors like CePt$_3$Si in the presence of 
interband pair scattering potential. We predict that apart from the
conventional transition temperature $T_c$, there is another cross-over 
temperature $T^*$ at which spin structure of the order parameter changes.
The order parameter changes from mixed singlet-triplet phase at
lower temperatures to only triplet
(singlet) phase for repulsive (attractive) interband scattering
potential at higher temperatures. The temperature dependence of
superfluid density shows a kink at this
cross-over temperature. We also have estimated critical field near the
conventional transition temperature.

\section*{Acknowledgments}
One of us (SPM) thanks CSIR, Govt. of India for his research fellowship.

\end{document}